\documentclass[journal]{IEEEtran}

\usepackage{slashbox}
\usepackage{graphicx}
\usepackage{subfigure}
\usepackage[justification=centering]{caption}       % Include this line if your
  \usepackage{amssymb}
 \usepackage{amsmath}
\usepackage{mathptmx}
\usepackage{times}
\usepackage{epstopdf}
\usepackage{fancyhdr}
\usepackage{subfigure}
 \usepackage{multirow}
\usepackage{color}
\usepackage{amsthm,amsmath,amssymb}
\usepackage{mathrsfs}
\begin{document}

\title{A review of Quantum Neural Networks: Methods, Models, Dilemma}
%An Open and Portable for Control Engineering
%Related Courses

\author{
	\IEEEauthorblockN{Renxin Zhao$^{1,2}$, Shi Wang$^{1}$} \\
	\IEEEauthorblockA{$^1$  College of Electrical and Information Engineering, Hunan University, Changsha, China}\\
	\IEEEauthorblockA{$^2$ School of Mechatronics and Automotive Engineering, Huzhou Vocational and Technical College, Huzhou, China}\\

    \thanks{Renxin Zhao is with Huzhou Vocational $\&$ Technical College, College of Mechanical and Electrical Engineering,
	No. 299 Xuefu Road, Huzhou, 313000, China. (e-mail:13061508@hdu.edu.cn) }
    \thanks{Shi Wang is with College of Electrical and Information Engineering, Hunan University, Changsha, 410082, China. (e-mail: peoplews3@hotmail.com)}
    \thanks{Corresponding author:  Shi Wang.}
}

\maketitle
\thispagestyle{fancy}
\fancyhead{}
\lhead{\footnotesize{TE-2020-000227}}
\rhead{\scriptsize{\thepage}}

\begin{table}[]
	\centering
	\begin{tabular}{ll}
		\multicolumn{2}{l}{Abbreviation}                                                          \\ \hline
		\multicolumn{1}{|l|}{CNN}   & \multicolumn{1}{l|}{Classical Neural Network}               \\ \hline
		\multicolumn{1}{|l|}{QNN}   & \multicolumn{1}{l|}{Quantum Neural Network}                 \\ \hline
		\multicolumn{1}{|l|}{CELL}  & \multicolumn{1}{l|}{Quantum Cellular Neural Network}        \\ \hline
		\multicolumn{1}{|l|}{NISQ}  & \multicolumn{1}{l|}{Noisy Intermediate-Scale Quantum}       \\ \hline
		\multicolumn{1}{|l|}{VQA}   & \multicolumn{1}{l|}{Variational Quantum Algorithm}          \\ \hline
		\multicolumn{1}{|l|}{VQC}   & \multicolumn{1}{l|}{Variational Quantum Circuit}            \\ \hline
		\multicolumn{1}{|l|}{CV}    & \multicolumn{1}{l|}{Continuous-Variable architecture}       \\ \hline
		\multicolumn{1}{|l|}{QP}    & \multicolumn{1}{l|}{Quantum Perceptron}                     \\ \hline
		\multicolumn{1}{|l|}{SONN}  & \multicolumn{1}{l|}{Self-Organizing Neural Network}         \\ \hline
		\multicolumn{1}{|l|}{QSONN} & \multicolumn{1}{l|}{Quantum Self-Organizing Neural Network} \\ \hline
		\multicolumn{1}{|l|}{QCPNN} & \multicolumn{1}{l|}{Quantum Competitive Neural Networks}    \\ \hline
		\multicolumn{1}{|l|}{CVNN}  & \multicolumn{1}{l|}{Convolutional Neural Network}           \\ \hline
		\multicolumn{1}{|l|}{QCVNN} & \multicolumn{1}{l|}{Quantum Convolutional Neural Network}   \\ \hline
		\multicolumn{1}{|l|}{GAN}   & \multicolumn{1}{l|}{Generative Adversarial Network}         \\ \hline
		\multicolumn{1}{|l|}{QGAN}  & \multicolumn{1}{l|}{Quantum Generative Adversarial Network} \\ \hline
		\multicolumn{1}{|l|}{GNN}   & \multicolumn{1}{l|}{Graph Neural Network}                   \\ \hline
		\multicolumn{1}{|l|}{QGNN}  & \multicolumn{1}{l|}{Quantum Graph Neural Network}           \\ \hline
		\multicolumn{1}{|l|}{Qubit} & \multicolumn{1}{l|}{Quantum bit}                            \\ \hline
		\multicolumn{1}{|l|}{QCPNN} & \multicolumn{1}{l|}{Quantum Competitive Neural Networks}    \\ \hline
		\multicolumn{1}{|l|}{TNN}   & \multicolumn{1}{l|}{Tensor Neural Network}                  \\ \hline
		\multicolumn{1}{|l|}{QTNN}  & \multicolumn{1}{l|}{Quantum Tensor Neural Network}          \\ \hline
		\multicolumn{1}{|l|}{RNN}   & \multicolumn{1}{l|}{Recurrent Neural Network}               \\ \hline
		\multicolumn{1}{|l|}{QRNN}  & \multicolumn{1}{l|}{Quantum Recurrent Neural Network}       \\ \hline
		\multicolumn{1}{|l|}{BM}    & \multicolumn{1}{l|}{Boltzmann Machine}                      \\ \hline
		\multicolumn{1}{|l|}{QBM}   & \multicolumn{1}{l|}{Quantum Boltzmann Machine}              \\ \hline
		\multicolumn{1}{|l|}{QWLNN} & \multicolumn{1}{l|}{Quantum Weightless Neural Network}      \\ \hline
		\multicolumn{1}{|l|}{RUS}   & \multicolumn{1}{l|}{Repeat-Until-Success}                   \\ \hline
		\multicolumn{2}{l}{\begin{tabular}[c]{@{}l@{}}* By adding a lowercase letter s after the abbreviation to indicate their\\ plural form\end{tabular}}
	\end{tabular}
\end{table}

\begin{abstract} The rapid development of quantum computer hardware has laid the hardware foundation for the realization of QNN. Due to quantum properties, QNN shows higher storage capacity and computational efficiency compared to its classical counterparts. This article will review the development of QNN in the past six years from three parts: implementation methods, quantum circuit models, and difficulties faced. Among them, the first part, the implementation method, mainly refers to some underlying algorithms and theoretical frameworks for constructing QNN models, such as VQA. The second part introduces several quantum circuit models of QNN, including QBM, QCVNN and so on. The third part describes some of the main difficult problems currently encountered. In short, this field is still in the exploratory stage, full of magic and practical significance.
\end{abstract}


\IEEEpeerreviewmaketitle

\section{Introduction}\label{sec0}
\IEEEPARstart{S}{ince} Feynman first proposes the concept of quantum computers \cite{1}, technology giants and startups such as Google, IBM, and Microsoft have competed with each other, eager to make it a reality. In 2019, IBM launches a quantum processor with 53 qubits, which can be programmed by external researchers. In the same year, Google announces that its 53-bit chip called Sycamore has successfully implemented "Quantum Supremacy". According to report, Sycamore can complete the world's fastest supercomputer IBM Summit in 200 seconds to complete calculations that take 10,000 years to complete, that is, quantum computers can complete tasks that are almost impossible on traditional computers \cite{2}. However, this statement is quickly doubted by competitors including IBM. IBM bluntly says: according to the strict definition of Quantum Supremacy, which means surpassing the computing power of all traditional computers, Google's goal of achieving Quantum Supremacy has not been achieved. Therefore, IBM issues an article criticizing Google's claim that traditional computers take 10,000 years to complete is wrong \cite{3}. Since IBM found that it only takes 2.5 days after the deduction, it also commented that Google has intensified the excessive hype about the current state of quantum technology \cite{3}. There are still nearly ten years left until 2030, which is called the first year of commercial use of quantum computing, Quantum computers that can be produced at this stage are all within the scope defined by NISQ. NISQ refers to the fact that there are fewer qubits available on quantum processors recently, and quantum control is susceptible to noise, but it already has stable computing power and the ability to suppress decoherence \cite{4}. In short, quantum computers are the hardware foundation for the development of QNN.

QNN is first proposed by \cite{8} and has been widely used in image processing \cite{9}-\cite{11}, speech recognition \cite{12}\cite{13}, disease prediction \cite{14}\cite{15} and other fields. For the definition of QNN, there is no unified conclusion in the academic circles. QNN is a promising computing paradigm that combines quantum computing and neuroscience \cite{16}. Specifically, QNN establishes a connection between the neural network model and quantum theory through the analogy between the two-level qubits, the basic unit of quantum computing, and the active/resting states in the complex signal transmission process in nerve cells \cite{17}. At current stage, it can also be defined as a sub-category of VQA, consisting of quantum circuits containing parameterized gate operations \cite{18}\cite{19}. Obviously, the definition of QNN can be completely different according to different construction methods \cite{20}-\cite{25}. In order to further clarify the precise meaning of QNN, \cite{26} puts forward the following three preconditions: (1) The initial state of the system can encode any binary string; (2) The calculation process can reflect the calculation principle of the neural network; (3) The evolution of the system is based on quantum effects and conforms to the basic theories of quantum mechanics. However, most of the QNNs models currently proposed are discussed on the level of mathematical calculations, and there are problems such as unclear physical feasibility, not following the evolution of quantum effects, and not having the characteristics of neural network computing. As a result, the real QNNs have not been realized \cite{26}.

From the perspective of historical development, QNN has roughly gone through two main stages, namely the early stage and the near-term quantum processor stage. In the early days, QNN could not be implemented on quantum computers due to hardware conditions. Most models were proposed based on the related physical processes of quantum computing, and did not describe specific quantum processor structures such as qubits and quantum circuits. Typical representatives are QNN based on multiple world views \cite{27}, QNN based on interactive quantum dots \cite{28}, QNN based on simple dissipative quantum gates \cite{29}, QNN analogous to CNN \cite{30}, and so on. Compared with earlier research results, the recently proposed QNN has a broader meaning. The term QNN is more often used to refer to a computational model with a networked structure and trainable parameters that is realized by a quantum circuit or quantum system \cite{31}. In addition, the research of the QNN model also emphasizes the physical feasibility of the model. In the recent quantum processor stage, some emerging models such as QBM \cite{32}-\cite{34}, QCVNN \cite{35}-\cite{37}, QGAN \cite{38}\cite{39}, QGNN \cite{40}-\cite{42}, QRNN \cite{43}\cite{44}, QTNN \cite{45}, QP\cite{46}-\cite{52}, etc. \cite{53}-\cite{63} will be introduced in subsequent sections.

QNN has surprising quantum advantages. But at this stage, the contradiction between dissipative dynamics in neural computing and unitary dynamics in quantum computing is still a topic worthy of in-depth study \cite{64}-\cite{70}. Furthermore, the current QNN can only be trained for some small samples at low latitudes, and the prediction accuracy and generalization performance in large data sets is still an open problem \cite{71}\cite{72}. In addition, it is also found that the barren plateau phenomenon is easily formed in the parameter space of the QNN exponential level \cite{73}-\cite{76}.

Finally, the main work of this article is summarized. In Section \ref{sec2}, the composition method of QNN will be introduced, so that readers have a preliminary understanding of the formation of QNN. In Section \ref{sec3}, the QNN quantum circuit model for the past six years will be introduced. In Section \ref{sec4}, some open issues and related attempts will be introduced.

\section{Some Construction Methods of QNN}\label{sec2}
%As discussed in Section \ref{sec0}, the purpose of the microcomputer control teaching experiment platform  is to facilitate researchers, students and engineers in acquiring the
Many related reviews are very enthusiastic about the QNN model, but they do not systematically tell us how to build a QNN. This will be an interesting topic. In fact, it is extremely difficult to systematically summarize all the methods from 1995 to 2021, so this section mainly reviews the relatively mainstream methods in the past 6 years.

\subsection{VQA}

VQC is a rotating quantum gate circuit with free parameters, which is used to approximate, optimize, and classify various numerical tasks. The algorithm based on VQC is called VQA, which is a classical-quantum hybrid algorithm, because the parameter optimization process usually takes place on classical computers \cite{18}.

The similarity between VQC and CNN is that they both approximate the objective function by learning parameters, but VQC has quantum characteristics. That is, all quantum gate operations are reversible linear operations, and quantum circuits use entanglement layers instead of activation functions to achieve multilayer structures. Therefore, VQC has been used to replace the existing CNN \cite{19}\cite{36}\cite{44}.

\begin{figure}[htbp]
	\vspace{0em}\centering
	\includegraphics[width=\linewidth]{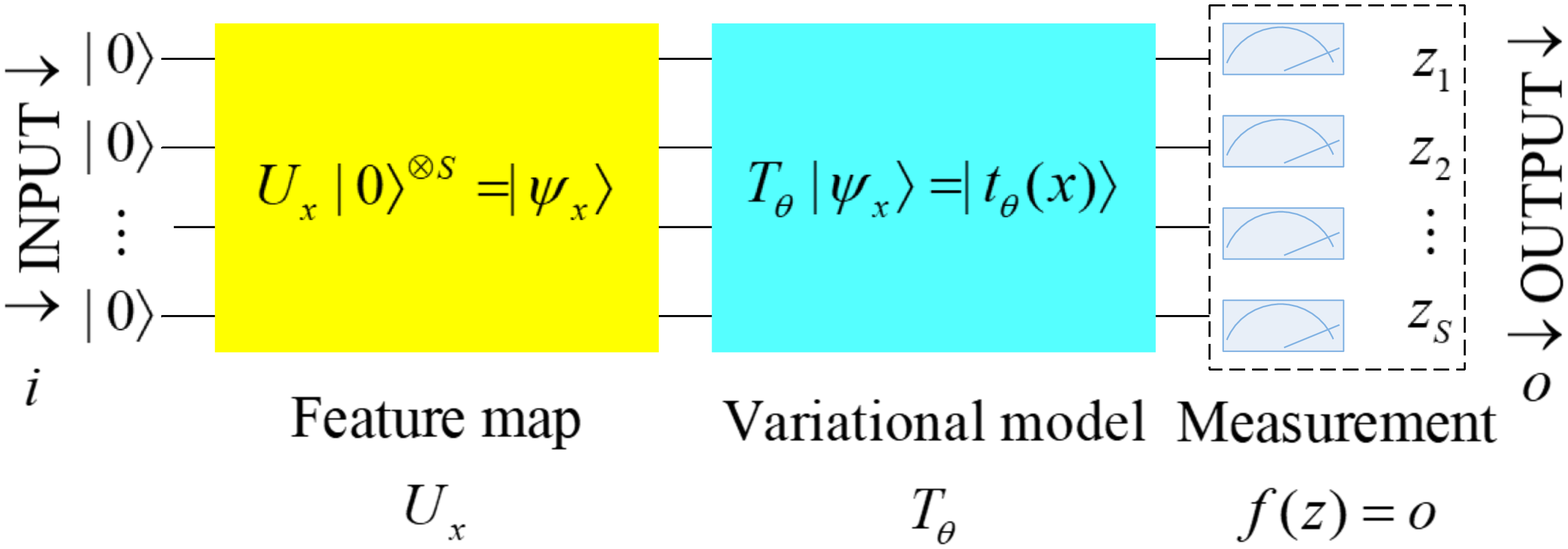}
	\vspace{0em}\caption{QNN based on VQA framework modified from \cite{19}.}\label{fig_1}
\end{figure}

A typical example is that, \cite{19} defines QNN as a subset of VQA and gives a general expression of QNN quantum circuit (see Fig. \ref{fig_1}). 

In Fig. \ref{fig_1}, in the first step, the quantum feature map method $|\psi_{x}\rangle:=U_{x}|0\rangle^{\otimes S}$ encodes the input information $i\in\mathbb{R}^{S_{i n}}$ which is usually in the form of classical data, into the \emph{S}-qubit Hilbert space. This step successfully realizes the transition from classical state to quantum state. Subsequently, the VQA containing parameterized gate operations optimized for specific tasks will play a role to evolve the obtained quantum state into a new state, namely $|t_{\theta}(x) \rangle:=T_{\theta}|\psi_{x}\rangle$, which is similar to the classical machine learning process \cite{19}. After the effect of VQA, the final output $o:=f(z)$ of QNN is extracted by quantum measurement. Before sending the information to the loss function, the measurement results $z= (z_{0},\ldots,z_{S})$ are usually converted into corresponding labels or predictions through classic post-processing. The purpose of this is to filter the parameters $\theta\in\Theta$ that can minimize the loss function for VQA.

The VQA framework is one of the mainstream methods for designing QNN. But it also inevitably inherited some of VQA's own shortcomings. For example, the QNN framework proposed by \cite{19}, in some cases, is facing the crisis of barren plateau. However, \cite{19} does not give specific solutions. Additionally, \cite{19} does not investigate the measurement efficiency of quantum circuits. Therefore, the QNN design under the VQA framework is still worth exploring.

\subsection{CV}

\begin{figure}[htbp]
	\vspace{0em}\centering
	\includegraphics[width=\linewidth]{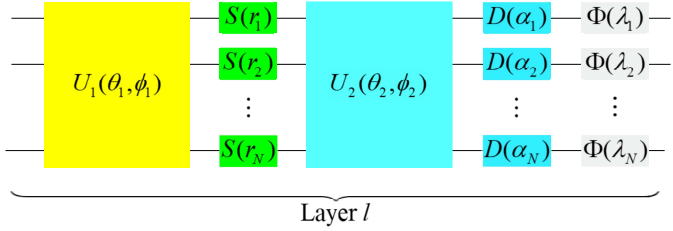}
	\vspace{0em}\caption{A single layer QNN based on CV framework modified from \cite{31}.}\label{fig_2}
\end{figure}

The idea of CV comes from \cite{31}.CV is a method for encoding quantum information with continuous degrees of freedom, and its specific form is VQC with a hierarchical structure including continuous parameterized gates. This structure has two outstanding points, namely the affine transformation realized by the Gaussian gate and the nonlinear activation function realized by the non-Gaussian gate. Based on the special structure of the CV framework, highly non-linear transformations can be encoded while retaining complete unity. The QNN framework based on CV is shown in Fig. \ref{fig_2}.

Fig. \ref{fig_2} shows the \emph{l}-th layer of the QNN model based on the CV framework. In Fig. \ref{fig_2} the universal \emph{N}-port linear optical interferometers $U_{i}=U_{i}(\theta_{i},\phi_{i})$  contain rotation gates as well as beamsplitter. In this figure, $S=\otimes_{i=1}^{N} S(r_{i})$ represents squeeze operators, collective displacement is marked by $D=\otimes_{i=1}^{N} D(\alpha_{i})$, and some non-Gaussian gates such as cubic phase or Kerr gates are represented by the symbol $\Phi=\Phi(\lambda)$. $(\theta,\phi,r,\alpha,\lambda)$ are collective gate variables and free parameters in the network, in which $\lambda$ can have a fixed value. The first interferometer $U_{1}$, the local squeeze gate \emph{S}, the second interferometer $U_{2}$ and the local displacement \emph{D} are used for affine transformation, and the last local non-Gaussian gate $\Phi$ is used for the final nonlinear transformation.

Being able to handle continuous variables is the bright spot of the CV-based QNN model, but one difficulty is how to realize the non-Gaussian gate, and to ensure that it has sufficient certainty and tunability. In this regard, \cite{31} does not give any further explanation. Moreover, \cite{31} has only done numerical experiments, and there is no practical application case yet.

\subsection{Swap Test and Phase Estimation}

\cite{20} and \cite{21} both suggest the method of swap test and phase estimation to build QNN. In the implementation scheme of \cite{20}, a single qubit controls entire input information of the neuron during the swap test, which is not conducive to the physical realization. Unlike \cite{20}, the quantum neuron in \cite{21} adopts the design of multi-phase estimation and multi-qubit SWAP test.

\begin{figure}[htbp]
	\vspace{0em}\centering
	\includegraphics[scale=0.38]{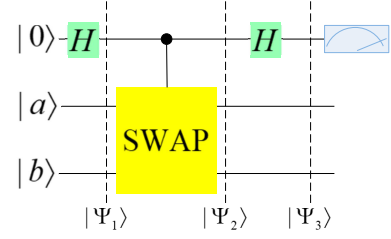}
	\vspace{0em}\caption{Quantum circuit of a swap test modified from \cite{21}.}\label{fig_3}
\end{figure}

\textbf{\emph{Swap Test}} \cite{22} first proposes swap test (see Fig. \ref{fig_3}). 

The meaning of this circuit is to know the square of the inner product $|\langle a \mid b\rangle|^{2}$ of the qubits $|a\rangle$ and $|b\rangle$, by measuring the probability that the first qubit is in the state $|0\rangle$ or $|1\rangle$. Fig. \ref{fig_3} has two Hadamard gates and a controlled swap operator. Assume that all states change to $|\Psi_{1} \rangle$ after going through the first Hadamard gate. Subsequently, $|\Psi_{1} \rangle$ is further transformed into $|\Psi_{2} \rangle$ under the action of the controlled swap operator. Finally, apply the Hadamard gate again, and $|\Psi_{3}\rangle$ is obtained. Performing projection measurement on ancilla qubits, we can know that the probabilities of $|0\rangle$ and $|1\rangle$ are $\frac{1}{2}+\frac{1}{2}|\langle a \mid b \rangle|^{2}$ and $\frac{1}{2}-\frac{1}{2}|\langle a \mid b \rangle|^{2}$, respectively.
Therefore, the square of the inner product of qubits $|a\rangle$ and $|b\rangle$ can be expressed as $|\langle a \mid b\rangle|^{2}=1-2 P(|1\rangle)$, where $P(|1\rangle)$ represents the probability of when the ancilla qubit is in state $|1\rangle$.

\begin{figure}[htbp]
	\vspace{0em}\centering
	\includegraphics[scale=0.5]{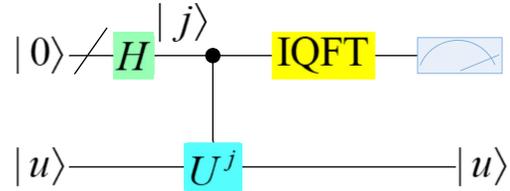}
	\vspace{0em}\caption{Quantum phase estimation modified from \cite{21}.}\label{fig_4}
\end{figure}

\textbf{\emph{Phase Estimation}} Assuming that $|u\rangle$ is an eigenvector of the unitary operator \emph{U}, the corresponding eigenvalue is $e^{2i\pi\varphi}$ and $\varphi$ is undetermined. Our goal is to obtain the estimated value of $\varphi$  through the phase estimation algorithm. As can be seen from Fig. \ref{fig_4}, there are two quantum registers.

The first one contains \emph{t} initial qubits with all states $|0\rangle$, and the second one starts in state $|u\rangle$. The phase estimation algorithm is implemented in three steps. In the first step, the circuit first applies the Hadamard transformation to the first register, and then applies a controlled-\emph{U} gates to the second register, where \emph{U} is raised to successive powers of two. The second step is to apply the inverse quantum Fourier transform represented by IQFT in Fig. \ref{fig_4} to the first register. The third step is to read the state of the first register by measuring on the basis of calculations.

\begin{figure*}[htbp]
	\vspace{0em}\centering
	\includegraphics[scale=0.35]{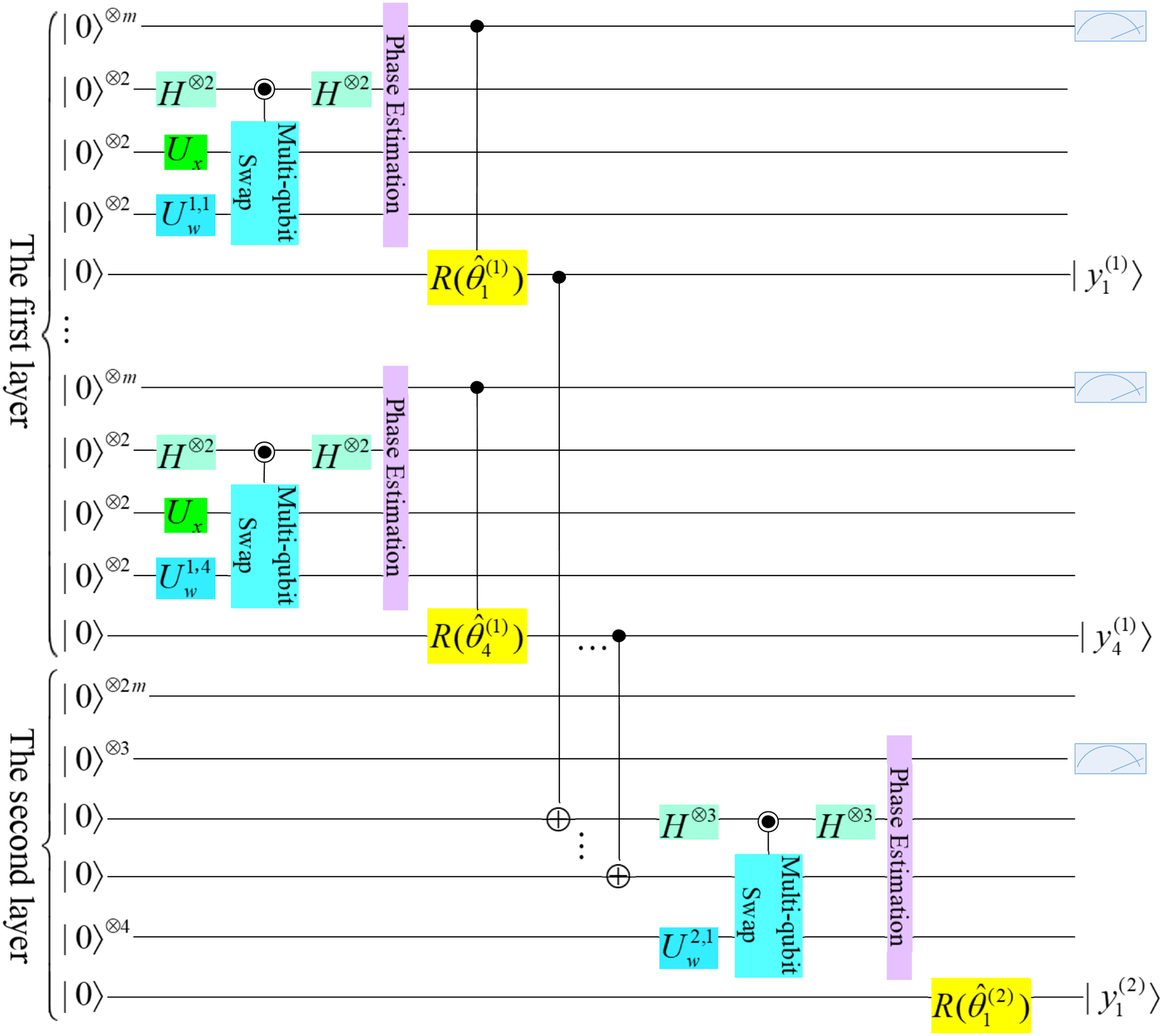}
	\vspace{0em}\caption{QNN based on swap test and phase estimation framework modified from \cite{21}.}\label{fig_5}
\end{figure*}

The QNN framework based on swap test and phase estimation proposed by \cite{21} is shown in Fig. \ref{fig_5}. This QNN framework converts the numerical sample into a quantum superposition state, and then obtains the inner product of the sample and the weight through the swap test, and then further maps the obtained inner product to the output of the quantum neuron through phase estimation. According to reports, since the framework does not need to record or store any measurement results, it will not waste classical computing resources. Although the model is more feasible to implement, in the case of multiple inputs, the input space will increase exponentially. Whether there will be a barren plateau is a question for further analysis.

\subsection{RUS}

\begin{figure}[htbp]
	\vspace{0em}\centering
	\includegraphics[scale=0.3]{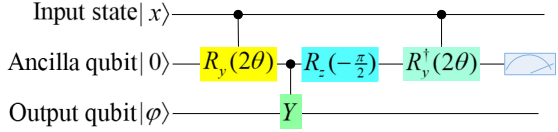}
	\vspace{0em}\caption{Repeat-until-success (RUS) circuit for realizing rotation with an angle $p(\phi)=\arctan(\tan^2\phi)$. \cite{23}.}\label{fig_6}
\end{figure}

\cite{23} uses the RUS quantum circuit \cite{24} to achieve an approximate threshold activation function on the phase of the qubit, and fully maintain the unitary characteristics of quantum evolution (see Fig. \ref{fig_6}).

A classic neuron can be seen as a function, which takes \emph{n} variables $x_{1}, \ldots, x_{n}$ and maps them to the output value $o=f (w_{1} x_{1}+\ldots+w_{n} x_{n}+b )$, where $ \{w_{i} \}$ and \emph{b} are synaptic weights and biases, respectively. $f(\cdot)$ is a non-linear activation function, such as step function, sigmoid function, tanh function, etc. We constrain the output value \emph{o} to be -1 and 1, that is $o \in [-1,1]$.

In order to map the above to the quantum framework, \cite{23} introduces some necessary quantum states:

1.$R_{y} (b \frac{\pi}{2}+\frac{\pi}{2} )|0\rangle=\cos  (b \frac{\pi}{4}+\frac{\pi}{4} )|0\rangle+\sin  (b \frac{\pi}{4}+\frac{\pi}{4} )|1\rangle$, where $b \in [-1,1]$ is a scalar;

2.$R_{y}(t)=e^{-itY/2}$ is a rotational quantum operation related to Pauli \emph{Y} operator. 

Extreme cases, such as $b=-1$ and $b=1$, will be mapped to quantum states $|0\rangle$ and $|1\rangle$, respectively, and $b \in (-1,1)$ will be regarded as a quantum neuron superimposed by $|0\rangle$ and $|1\rangle$. Take $|x\rangle=|x_{1} \ldots x_{n} \rangle$ as the control state, and utilize $R_{y}=(2w_{i})$ to the ancilla qubit conditional on the \emph{i}-th qubit, and then apply $R_{y}=(2b)$ to the ancilla qubit. This is equivalent to applying $R_{y}=(2\theta)$ to the ancilla qubit conditioned on the state $x_{i}$ of the input neuron. Rotation is performed by $R_{y}=(2f(\theta))$ \cite{23}.

Fig. \ref{fig_6} depicts a circuit that carries out $R_{y}=(2p(\phi))$, where $p(\phi)=\arctan(\tan^2\phi)$ is a nonlinear activation function. The measurement result of the ancilla qubit demonstrates the influence of the RUS circuit on the output qubit. The measurement returns to $|0\rangle$, denoting that the rotation of $2p^{ok}(\phi)$ is successfully achieved to the output qubit. On the contrary, if it is $|1\rangle$, $R_{y}(\pi/2)$ is rotated on the output qubit. At this time, $R_{y}(-\pi/2)$ needs to be used to offset this rotation. Then the circuit keeps repeating until $|0\rangle$ is detected on the ancilla qubit, which is why it is called RUS

The highlight of \cite{23} is to use quantum circuits to approximate nonlinear functions, that is, to solve nonlinear problems by linear means. This unifies nonlinear neural computing and linear quantum computing, and meets the basic requirements of \cite{26}. It is also worth mentioning that RUS is a flexible way. Quantum neurons constructed with RUS also have the potential to construct various machine learning paradigms, involving supervised learning, reinforcement learning and so on.

\subsection{Quantum generalization}

\cite{25} puts forward a quantum generalization method for CNN, that is, the reformation of the perceptron can be explained by each reversible and unitary transformation in QNN. Through the use of numerical simulations, it has been proven that gradient descent can be used to train for a given objective function. Minimizing the difference between the expected output and the output of the quantum circuit is the purpose of training. One feasible physical approach is to apply quantum photonics.

Although the theory of \cite{25} is universal, it ignores the nonlinear problem when discussing the quantum generalization of QNN.

\section{Advances of QNN Models for Near-term Quantum Processor}\label{sec3}

\subsection{QBM}

\begin{figure}[htbp]
	\vspace{0em}\centering
	\includegraphics[scale=0.4]{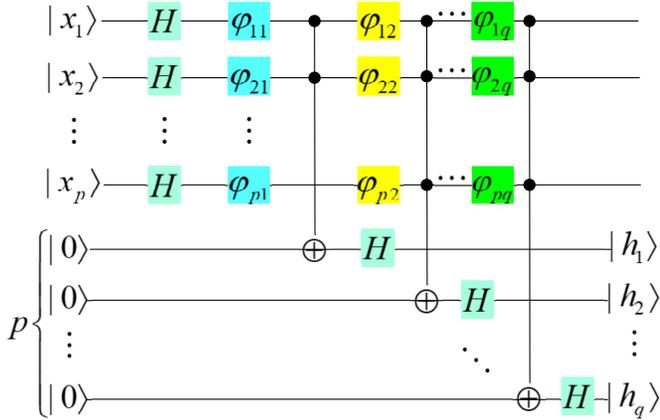}
	\vspace{0em}\caption{The quantum circuit of QRBM modified from \cite{32}.}\label{fig_7}
\end{figure}

\cite{32} provides a new idea for the realization of QBM. Adopting the input represented by the quantum state, employing quantum gates for data training and parameter updating, by modeling the quantum circuits of visible layers and hidden layers, the global optimal solution can be turned up by QRBM. (see Fig. \ref{fig_7}).

In Fig. \ref{fig_7}, the visible layer variable is expressed as $(|v_{1}\rangle,|v_{2}\rangle,\cdots,|v_{p}\rangle)$, where \begin{equation}
	\begin{aligned}
		|v_{i}\rangle=
		& \cos (\frac{2 \pi(v_{i}-\min _{1 \leq j \leq p}\{v_{j}\})}{\max _{1 \leq j \leq p}\{v_{j}\}-\min _{1 \leq j \leq p}\{v_{j}\}})|0\rangle \\
		&+\sin (\frac{2 \pi(v_{i}-\min _{1 \leq j \leq p}\{v_{j}\})}{\max _{1 \leq j \leq p}\{v_{j}\}-\min _{1 \leq j \leq p}\{v_{j}\}})|1\rangle
	\end{aligned}
\end{equation} and $i=1,\ldots,p$. The hidden layer is denoted as $( |h_{1}\rangle,|h_{2}\rangle,\cdots,|h_{p}\rangle)$, where \begin{equation}
\begin{aligned}
	&|h_{k} \rangle=\cos  (\frac{2 \pi (h_{k}-\min _{1 \leq I \leq q} \{h_{l} \} )}{\max _{1 \leq 1 \leq q} \{h_{l} \}-\min _{1 \leq I \leq q} \{h_{l} \}} )|0\rangle \\
	&+\sin  (\frac{2 \pi (h_{k}-\min _{1 \leq 1 \leq q} \{h_{l} \} )}{\max _{1 \leq 1 \leq q} \{h_{l} \}-\min _{1 \leq l \leq q} \{h_{l} \}} )|1\rangle
\end{aligned}
\end{equation} and $k=1, \ldots, q$. The quantum register has \emph{p} qubits. The Hadamard gate in Fig. \ref{fig_7} is used for preprocessing \cite{32}. The coefficients of the visible layer variables are changed with the phases of a series of quantum rotation gates. The quantum state of each variable is switched through the CNOT gate, and the entire variable in the visible layer is summed into one qubit \cite{32}. After passing through the Hadamard gate again, the quantum state of a qubit in the hidden layer is obtained to represent the output.

In recent years, QBM models based on variable quantum algorithm have also been proposed. \cite{33} proposes a variational virtual-time simulation based on NISQ equipment to realize BM learning. It is different from the previous method of preparing thermal equilibrium, but uses a pure state whose distribution simulates the thermal equilibrium distribution. It has been proved that NISQ equipment has the potential of effective use in BM learning. \cite{34} prepares the Gibbs state and evaluates the analytical gradient of the loss function based on the variable quantum virtual time evolution technology. Numerical simulations and experiments are carried out on IBM Q to prove the approximation of the variational Gibbs state. Compared with \cite{33} and \cite{34}, \cite{32} realized a pioneering effort to realize QBM with quantum gates, explored the appropriate number of hidden layers, and tested the pattern recognition performance of gearboxes with different hidden layers.

\subsection{QCVNN}

QCVNN has received extensive attention in the past three years. \cite{35} first mentiones the term QCVNN. In \cite{35}, the input information is represented by qubits, which are trained under the CVNN framework, and the probability of a certain characteristic state is obtained by measurement. But what \cite{35} puts forward is only a theoretical model, which does not have the feasibility of quantum circuits.

\begin{figure}[htbp]
	\vspace{0em}\centering
	\includegraphics[scale=0.43]{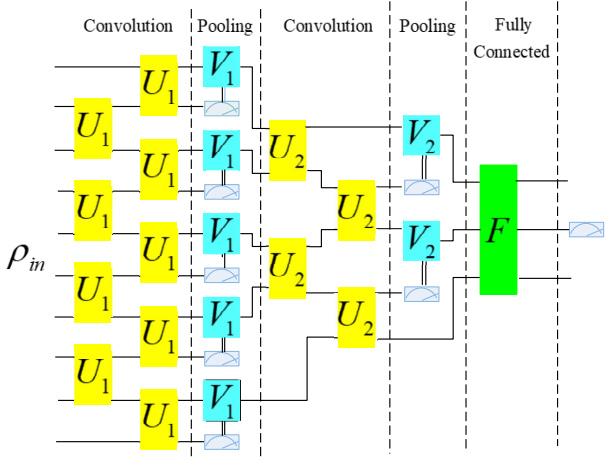}
	\vspace{0em}\caption{The structure of QCVNN modified from \cite{36}.}\label{fig_8}
\end{figure}

\cite{36} designs a quantum circuit model of QCVNN, which implements convolution and pooling transformation similar to CVNN for processing one-dimensional quantum data. The quantum circuit structure of QCVNN is shown in Fig. \ref{fig_8}, including several repeated convolutional layers and pooling layers, as well as a fully connected layer. 

The convolutional layer imposes a finite depth of quasi-local unitary transformation, and each unitary transformation parameter in the same convolutional layer is the same. The pooling layer measures some qubits and applies classical controlled unitary transformation on adjacent qubits. The parameters of the unitary transformation depend on the measurement results of adjacent qubits. After multi-layer convolution and pooling transformation, when there are few remaining qubits, a unitary transformation is applied as a fully connected layer, and the specified qubits are measured to obtain the judgment result of the network. Each convolutional layer and pooling layer of the network share the unitary transformation of the same parameter, so for \emph{n}-bit input qubits, it only has parameters of the order of $O(\log (n))$, which can be efficiently trained and deployed on quantum computers in near term. In addition, the pooling layer can reduce the dimensionality of the data and introduce a certain non-linearity through partial measurement. It is worth noting that as the convolutional layer increases, the number of qubits spanned by the convolution operation will also increase. Therefore, the deployment of the QCVNN model on a quantum computer requires the ability to implement two-qubit gates and projection measurements at various distances.

\begin{figure*}[htbp]
	\vspace{0em}\centering
	\includegraphics[scale=0.29]{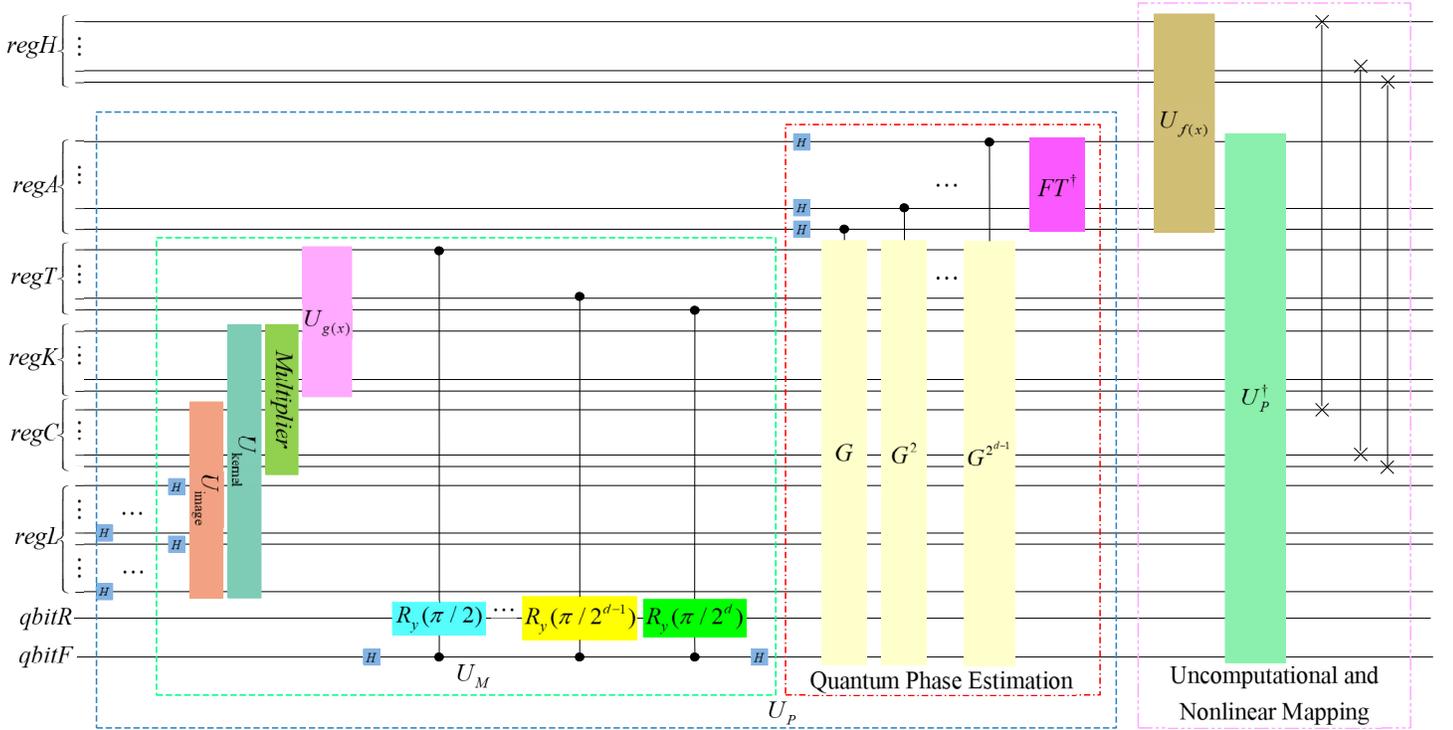}
	\vspace{0em}\caption{The quantum circuit for the convolutional layer in deep QCVNN model modified from \cite{37}.}\label{fig_9}
\end{figure*} 

\cite{37} introduces a deep QCVNN model for image recognition (see Fig. \ref{fig_9}). The general working principle of the model is that the input image will be encoded into quantum states with basis encoding, and then these quantum states undergo a series of parameter-related unitary transformations to realize the quantum evolution process. Unlike CVNN, this model omits the pooling layer, only retains the quantum convolution layer and quantum classification layer, and increases the convolution step size to $2^m$ to finish sub-sampling. This is advantageous to remove the intermediate measurement, so as to achieve the purpose of quickly reducing the dimension of the generated feature map. Finally, the corresponding classification label is acquired through quantum measurement. For some details, for example, in the quantum convolution layer, the quantum preparation of the input image and the related convolution kernels is performed by the QRAM algorithm. For another example, the quantum inner product operation of the kernel working area requires the support of a quantum multiplier and a controlled Hadamard gate rotation operation. Furthermore, the conversion between amplitude coding and base coding and nonlinear mapping are determined by the quantum phase estimation algorithm. Finally, by separating the desired state and the intermediate state, non-computational operations are employed to obtain the input state of the next layer. 

The proposal of the deep QCVNN model proves the feasibility and efficiency in multi-type image recognition. However, this model can only limit the size of the input image to the range of $2^{n} \times 2^{n}$. Once it is not in this range, the image needs to be scaled additionally. For the image scaling problem, \cite{37} does not give a good quantum version solution. Furthermore, the optional step size and nucleus still need to be further probed.

Comparing \cite{36} and \cite{37}, \cite{36} only gives a QCVNN model for low-dimensional input. Although \cite{36} also mentiones that the proposed model is extensible, it does not elaborate on the expansion steps. Relatively speaking, the input dimension discussed by deep QCVNN is higher.

\subsection{QGAN}

The concept of quantum generative adversarial learning can be traced back to \cite{38}. \cite{38} discourses the operating efficiency of quantum generative adversarial learning in a variety of situations, such as the training data is classical data or quantum data, and whether the discriminator and generator use quantum processors to run. Reportedly, when the training data is quantum data, the quantum confrontation network may show an exponential advantage over the classical confrontation network. However, \cite{38} does not give a specific quantum circuit scheme, nor does it conduct a rigorous mathematical derivation.

\begin{figure}[htbp]
	\vspace{0em}\centering
	\includegraphics[scale=0.45]{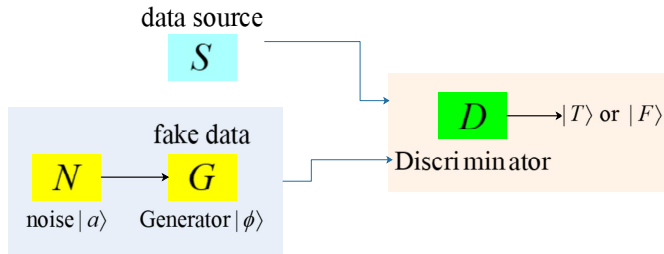}
	\vspace{0em}\caption{The schematic diagram of QGAN modified from \cite{39}.}\label{fig_10}
\end{figure}

\begin{figure}[htbp]
	\vspace{0em}\centering
	\includegraphics[scale=0.32]{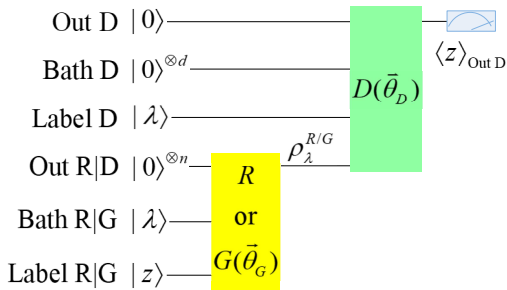}
	\vspace{0em}\caption{The general structure diagram of QGAN modified from \cite{39}.}\label{fig_11}
\end{figure}

A quantum circuit version of GAN is proposed by \cite{39}. The schematic diagram is shown in Fig. \ref{fig_10} and the structure diagram is shown in Fig. \ref{fig_11}.

\cite{39} assumes that there is a data source (\emph{S}) and a label $|\phi\rangle$ is given, and the density matrix $\rho_{\phi}^{S}$ is output to a register comprising \emph{n} subsystems, namely
\begin{equation}
	R(|\phi\rangle)=\rho_{\phi}^{S}
\end{equation}

In \cite{39}, the essence of the Generator (\emph{G}) is VQC, and its gate is parameterized by the vector $\vec{\theta}_{G}$. Regarding the label $|\phi\rangle$ and the additional state $|a\rangle$ as inputs, \emph{G} generates a quantum state
\begin{equation}
	G (\vec{\theta}_{G},|\phi, a\rangle )=\rho_{\phi}^{G} (\vec{\theta}_{G}, a )
\end{equation}

In (4), $\rho_{\phi}^{G}$ is output on a register containing \emph{n} subsystems, similar to the real data \cite{39}. The additional input state $|a\rangle$ has a dual role. On the one hand, it can be seen as an unstructured noise source that supplies entropy in the distribution of the generated data. On the other hand, it can be regarded as a control for \emph{G}.

The training signal of \emph{G} is given by the Discriminator (\emph{D}), which consists of an independent quantum circuit parameterized by the vector $\vec{\theta}_{D}$. The function of \emph{D} is to judge whether the given input state comes from \emph{S} or \emph{G}. During this period, \emph{G} will deceive \emph{D} continuously, and try its best to make \emph{D} believe that its output is True (\emph{T}, "Fake" denoted as \emph{F}). Assuming the input is from \emph{S}, the output register of \emph{D} will output $|T\rangle$, otherwise it will output $|F\rangle$. In the internal work area, \emph{D} can also perform operations. So as to cpompel \emph{G} to comply with the provided label, \emph{D} is also given a copy of the unchanged label.

Adversarial tasks can describe the optimization goals of QGAN
\begin{equation}
	\begin{aligned}
		\min _{\vec{\theta}_{G}} \max _{\bar{\theta}_{D}} & \frac{1}{\phi} \sum_{\phi=1}^{\phi} \operatorname{Pr}((D(\vec{\theta}_{D},|\phi\rangle, R(|\phi\rangle))=|T\rangle) \\
		&\cap(D(\vec{\theta}_{D},|\phi\rangle, G(\vec{\theta}_{G},|\phi, a\rangle))=|F\rangle))
	\end{aligned}
\end{equation}

After clarifying \emph{G}, \emph{D} and optimization goals, \cite{39} gives the general structure of QGAN, as shown in Fig. \ref{fig_11}. The initial states are respectively defined on the registers labeled \textbf{Label R$\mid$G}, \textbf{Out R$\mid$G} and \textbf{Bath R$\mid$G}, which can be applied by \emph{S} or the parametrized \emph{G} $G(\vec{\theta}_{G})$. The initial resource state $|0,0,\lambda\rangle$ defined in the \textbf{Out D}, \textbf{Bath D} and \textbf{Label D} registers and the information  $\rho_{\phi}^{S / G}$ from \emph{S} are available for \emph{D} $D(\vec{\theta}_{D})$ to use. \emph{D} announces whether the result is $|T\rangle$ or $|F\rangle$ in the \textbf{Out D} register. The expected value $\langle Z\rangle_{\text {Out } \mathrm{D}}$ is proportional to the probability that \emph{D} will output $|T\rangle$.

\cite{39} proved the feasibility of QGAN's explicit quantum circuit through a simple numerical experiment. QGAN has more extensive characterization capabilities than the classic version, for example, it can learn to generate encrypted data.

\subsection{QGNN}

A deep learning architecture that can directly learn end-to-end, Quantum Space Graph Convolutional Neural Network (QSGCNN), was proposed by \cite{40} to classify graphs of any size. The main idea is to transform each graph into a fixed-size vertex grid structure through the transfer alignment between graphs, and use the proposed quantum space graph convolution operation to propagate the vertex features of the grid. According to reports, the QSGCNN model not only retains the original map features, but also bridges the gap between the spatial map convolutional layer and the traditional convolutional neural network layer, and can better distinguish different structures \cite{40}.

Quantum Walking Neural Network (QWNN), a graph neural network structure based on quantum random walk, constructs diffusion operators by learning quantum walks on the graph, and then applies them to graph structure data \cite{41}. QWNN can adapt to the space of the entire graph or the time of walking. The final diffusion matrix can be cached after learning has converged, so that it can be quickly advanced in the network. However, due to the constant shift operation and coin insertion in the learning process, this model is significantly slower than other models. Space complexity is called a problem worthy of continued analysis.

\begin{figure}[htbp]
	\vspace{0em}\centering
	\includegraphics[scale=0.3]{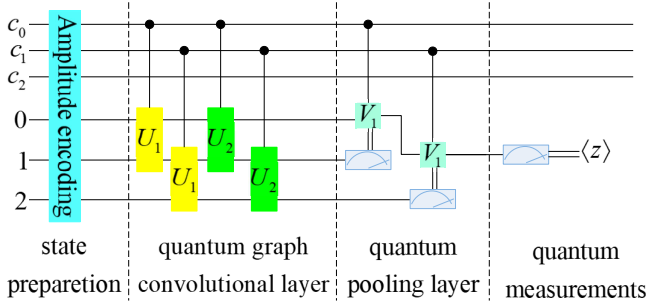}
	\vspace{0em}\caption{A QGCN quantum circuit model modified from \cite{42}.}\label{fig_12}
\end{figure}

The above models belong to the theoretical framework. The quantum circuit model of QGNN was studied by \cite{42}, as shown in Fig. \ref{fig_12}.

Quantum state preparation, quantum graph convolutional layer, quantum pooling layer and quantum measurements constitute the quantum circuit model of Fig. \ref{fig_12}. In the state preparation stage, the image data is effectively encoded into a quantum state by the amplitude encoding method. The normalized classical vector $x \in \mathbb{C}^{2^{n}}, \sum_{j} |x_{j} |^{2}=1$ can be represented by the quantum state $|\psi\rangle$ as follows:
\begin{equation}
	x= (x_{1}, \ldots, x_{2^{n}} )^{\mathrm{T}}  \rightarrow |\psi_{x} \rangle=\sum_{j=1}^{2^{n}} x_{j}|j\rangle
\end{equation}

In the same way, a classic matrix $B \in \mathbb{C}^{2^{n} \times 2^{m}}$ with $a_{ij}$ that satisfies $\sum_{i j} |a_{i j} |^{2}=1$ can be encoded as $|\psi\rangle=\sum_{i=1}^{2^{m}} \sum_{j=1}^{2^{n}} a_{i j}|i\rangle|j\rangle$ by expanding the Hilbert space accordingly. In the quantum graph convolutional layer, the constructed dual-qubit unitary operation \emph{U} realizes local connection. The number of layers of the quantum graph convolutional layer indicates the order of node aggregation, so the unitary operations of the same layer have the same parameters, which reflects the characteristics of parameter sharing. In the quantum pooling layer, quantum measurement is added to reduce the feature dimension, achieving the same effect as the classical pooling layer. Note, however, that not all qubits are measured, but a part of it is measured. Based on the measurement results, it is determined whether to perform unitary transformation on adjacent qubits. Finally, after multi-layer convolution and pooling transformation, the specified qubit can be measured by quantum, and the expected value can be obtained. The results show that the structure can effectively capture node connectivity and learn the hidden layer representation of node features.

The model proposed by \cite{42} can effectively deal with the problem of graph-level classification. The four major structures can effectively capture the connectivity of nodes, but currently only the node information is used, and the characteristics of edges are not studied.

\subsection{QRNN}

\begin{figure}[htbp]
	\vspace{0em}\centering
	\includegraphics[scale=0.4]{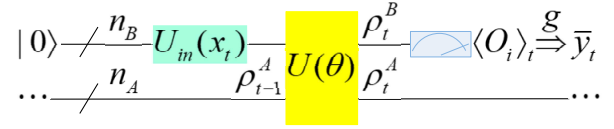}
	\vspace{0em}\caption{Structure of QRNN for a single time step modified from \cite{43}.}\label{fig_13}
\end{figure}

\cite{43} constructs a parameterized quantum circuit similar to the RNN structure. Some qubits in the circuit are used to memorize past data, while other qubits are measured and initialized at each time step to obtain predictions and encode a new input data. The specific structure is shown in Fig. \ref{fig_13}.

In Fig. \ref{fig_13}, there are two groups of qubits, denoted by $n_{A}$ and $n_{B}$ respectively. The qubits of group A have never been measured, and they are used to retain past information. The qubits of group B are measured every time step \emph{t}, and initialized at the same time to output prediction and input to the system. The time step here is composed of three parts, namely the encoding part, the evolution part and the measurement part. In the encoding part, $U_{in}(x_{t})$ is used for the initial state $|0\rangle$ (abbreviate $|0\rangle^{\otimes n_{B}}$ as $|0\rangle$), and the training data $x_{t}$ is encoded into the quantum state of the B group of qubits. The information about $x_{0},\ldots,x_{t-1}$ has been saved in group A as the density matrix $\rho_{t-1}^{A}= (\theta, x_{0}, \ldots, x_{t-1} )$ generated in the previous step. In the evolution part, the parameterized unitary operator $U(\theta)$ acts on the entire qubit, which can transfer information from group B to group A. Here, use the evolved $\rho_{t}^{A}$ and $\rho_{t}^{B}$ to represent the simplified density matrices of A and B, respectively. In the measurement part, first measure the expected value of a group of observations $\{O_{i}\}$ in group B to obtain 
\begin{equation}
	 \langle O_{i} \rangle_{t}=\operatorname{Tr} [\rho_{t}^{B} O_{i} ]
\end{equation}

Then, the expected value is transformed into a certain function g, and the prediction $\overline{y_{t}}=g(\{\langle O_{i}\rangle_{t}\})$ of $y_{t}$ is obtained. \cite{43} points out that the transformation g can be chosen arbitrarily, for example, g can be a linear combination of $\{\langle O_{i}\rangle_{t}\}$. Finally, the qubits in group B are initialized to $|0\rangle$.  After repeating these three parts many times,  $y_{0}, \ldots, y_{T-1}$'s prediction $\overline{y_{0}}, \ldots, \overline{y_{T-1}}$ can be obtained. After the prediction is obtained, the cost function L is calculated, which represents the difference between the training data $\{y_{0}, \ldots, y_{T-1}\}$ and the predicted data $\{\overline{y_{0}}, \ldots, \overline{y_{T-1}}\}$ obtained by QRNN. The parameter $\theta$ is optimized by a standard optimizer running on a classic computer to minimize L.

The QRNN model proposed by \cite{43} is a parameterized quantum circuit with a recursive structure. The performance of the circuit is determined by three parts: (1) different data encoding units (2) the structure of the parameterized quantum circuit (3) the optimizer used to train the circuit. For the first point, a simple door is used as a demonstration in the article. For the second point, we can still explore further. For the third factor, we can learn from the method of VQA. But an unresolved question is whether QRNN is better than the classic one. This problem requires the establishment of some indicators for further analysis and experimentation.

\subsection{QTNN}

\begin{figure}[htbp]
	\vspace{0em}\centering
	\includegraphics[scale=0.5]{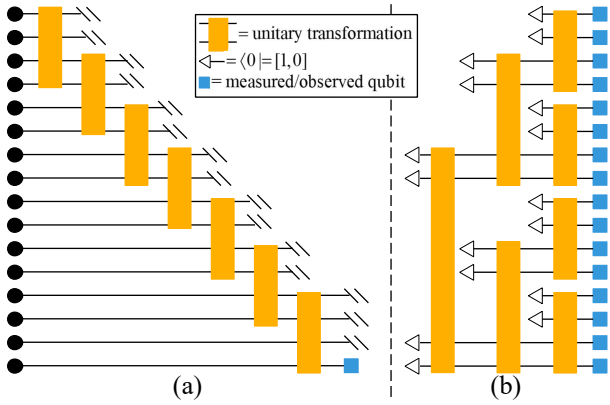}
	\vspace{0em}\caption{Quantum tensor networks modified from \cite{45}. (a) The discriminative network; (b) the generative network.}\label{fig_14}
\end{figure}

\cite{45} is the first to explore quantum tensor neural networks for recent quantum processors (see Fig. \ref{fig_14}).

The author proposes a QNN with a tensor network structure, which can be used for discrimination tasks and generation tasks. The neural network model has a fixed tree structure quantum circuit, in which the parameters of the unitary transformation are not fixed initially, and the training algorithm is a quantum-classical hybrid algorithm, that is, searching for suitable parameters with the aid of a classical algorithm.

In Fig. \ref{fig_14} (a), the discriminant model encodes the input data $\mathbf{x}=(x_{1}, \cdots x_{n}), 0 \leq x_{i} \leq 1$ as a product state of \emph{n} qubits, as the input quantum state
\begin{equation}
	|\mathbf{x}\rangle= (\begin{array}{c}
		\cos  (\frac{\pi}{2} x_{1} ) \\
		\sin  (\frac{\pi}{2} x_{1} )
	\end{array} ) \otimes (\begin{array}{c}
		\cos  (\frac{\pi}{2} x_{2} ) \\
		\sin  (\frac{\pi}{2} x_{2} )
	\end{array} ) \otimes \cdots \otimes(\begin{array}{c}
		\cos  (\frac{\pi}{2} x_{n} ) \\
		\sin  (\frac{\pi}{2} x_{n} )
	\end{array})
\end{equation}

The quantum circuit of the discriminant model is similar to a multilayer tree structure. After applying the unitary transformation, half of the qubits are ignored, and the remaining qubits continue to participate in the operation of the next layer of nodes. Finally, one or more qubits are used as output qubits, and the most probable measurement result is used as the judgment result of the input data by the neural network. In the training process, the classical algorithm is used to compare the discrimination result with the real label, and the circuit parameters are updated according to the error.

The generative model adopts a structure almost opposite to the discriminant model, as shown in Fig. \ref{fig_14} (b). The newly added qubit is combined with the original qubit to participate in the calculation of the lower unitary transformation node. After  generating the required number of qubits, the data information generated by the QNN model is obtained through measurement. When training the network, the quantum circuit parameters are also adjusted through classical algorithms according to the generated results and label errors.

The tensor network provides an increasingly complex natural hierarchical structure of quantum states, which can reduce the number of qubits required to process high-dimensional data with the support of dedicated algorithms. In addition, it can alleviate the problems related to random initial parameters, and it seems to have application potential in the noise recovery ability of machine learning algorithms. In addition, tensor network is a very promising framework because it has achieved a careful balance between expressive power and computational efficiency, and has rich theoretical understanding and powerful optimization algorithms.

\subsection{QP}

QP is one of the relatively mature models. The smallest organ of QP is a quantum neuron \cite{23}. \cite{46} establishes a model of neuron and concludes: a single quantum neuron can perform an XOR function that cannot be achieved by a classical neuron, and has the same computing power as a double-layer perceptron. Quantum neurons also have variants such as feedback quantum neurons \cite{47} and artificial spike quantum neurons \cite{48}.

In view of the core position of neurons in the multilayer perceptron, \cite{49} proposes an artificial neuron that can be carried out an real quantum processor. The circuit it provides is as follows.
\begin{figure}[htbp]
	\vspace{0em}\centering
	\includegraphics[scale=0.4]{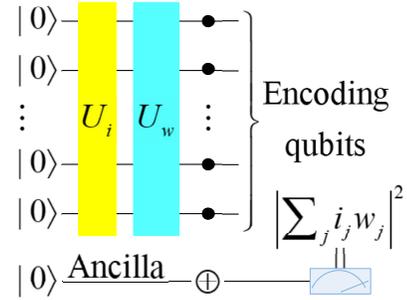}
	\vspace{0em}\caption{Scheme of the quantum algorithm for the implementation of the artificial neuron model on a quantum processor, modified from \cite{49} }\label{fig_15}
\end{figure} 

Fig. \ref{fig_15} outlines the principle. The system starts from the state $|00\ldots\rangle$, and after passing through the unitary matrix $|\psi_{i}\rangle$, $|00\ldots\rangle$ is transformed into the input quantum state $|\psi_{i}\rangle$. $|\psi_{i}\rangle$ undergoes transformation $U_{w}$, transfers the result to an ancilla qubit, and finally performs quantum measurement on it to assess the activation state of the perceptron.

More specifically, the input vector and weight are restricted to the binary value $i_{j},w_{j} \in \{-1,1\}$. Given any input $\overline{i}$ and weight $\overline{w}$ vector, use \emph{m} coefficients required to define the general wave function $|\psi_{i}\rangle$ of \emph{N} qubits to encode the \emph{m}-dimensional input vector.
\begin{equation}
	\begin{aligned}
		&\bar{i}= (i_{0}, \ldots, i_{m-1} )^{\mathrm{T}} \\
		&\bar{w}= (w_{0}, \ldots, w_{m-1} ight)^{\mathrm{T}} \quad \text { with } i_{j}, w_{j} \in\{-1,1\}
	\end{aligned}
\end{equation}

Next, two quantum states are defined
\begin{equation}
	|\psi_{i}\rangle=\frac{1}{\sqrt{m}} \sum_{j=0}^{m-1} i_{j}|j\rangle,|\psi_{w}\rangle=\frac{1}{\sqrt{m}} \sum_{j=0}^{m-1} w_{j}|j\rangle
\end{equation} where $|j\rangle \in\{|00 \ldots 00\rangle,|00 \ldots 01\rangle, \ldots,|11 \ldots 11\rangle\}$. As described in (10), the factor $\pm 1$ can be used to encode the \emph{m}-dimensional classical vector as a uniform weighted superposition of the complete calculation basis $|j\rangle$.
 
First, encode the input value in $\overline{i}$ to prepare the $|\psi_{i}\rangle$ state. Assuming that the initial state of the qubit is $|00 \cdots 00\rangle \equiv |0\rangle^{\otimes N}$, a unitary transformation $U_{i}$ is performed such that
\begin{equation}
	U_{i}|0\rangle^{\otimes N}= |\psi_{i}\rangle
\end{equation}

In principle, any $m \times m$ unitary matrix with $\overline{i}$ in the first column can be used for this purpose. In a more general case, starting from a blank register to prepare the input state may be replaced by the quantum memory stored before $|\psi_{i}\rangle$ is directly called.

In the second step, the quantum register is used to calculate the inner product between $\overline{i}$ and $\overline{w}$. By defining the unitary transformation $U_{w}$ and rotating the weighted quantum state to
\begin{equation}
	U_{w}|\psi_{w}\rangle=|1\rangle^{\otimes N}=|m-1\rangle
\end{equation}, the task can be performed effectively. Any $m \times m$ unitary matrix with $\bar{w}^{\mathrm{T}}$ in the last row meets this condition. If $U_{w}$ is added after $U_{i}$, the overall \emph{N}-qubit quantum state becomes
\begin{equation}
	U_{w}|\psi_{i}\rangle=\sum_{j=0}^{m-1} c_{j}|j\rangle \equiv |\phi_{i, w}\rangle
\end{equation}

According to (12), the scalar product between two quantum states is
\begin{equation}
	\begin{gathered}
		\langle\psi_{w} \mid \psi_{i}\rangle=\langle\psi_{w}|U_{w}^{\dagger} U_{w}| \psi_{i}\rangle \\
		=\langle m-1 \mid \phi_{i, w}\rangle=c_{m-1}
	\end{gathered}
\end{equation}

According to the definition of (10), the scalar product of the input vector and the weight vector is $\bar{w}, \bar{i}=m\langle\psi_{w} \mid \psi_{i}\rangle$. Therefore, the desired result is contained in the coefficient $c_{m-1}$ of the final state $|\phi_{i, w}\rangle$, which reaches a normalization factor.

To extract such information, an ancilla qubit (\emph{a}) initially set to state $|0\rangle$ is used. There is a multi-control NOT gate between the \emph{N}-coded qubit and target \emph{a} leading to
\begin{equation}
	|\phi_{i, w}\rangle|0\rangle_{a} \rightarrow \sum_{j=0}^{m-2} c_{j}|j\rangle|0\rangle_{a}+c_{m-1}|m-1\rangle|1\rangle_{a}
\end{equation}

The required nonlinearity of the threshold function of the perceptron output is obtained immediately by performing a quantum measurement. Measuring the state of ancilla qubits on the basis of the calculation produces output $|0\rangle_{a}$ (i.e., an active perceptron) with probability $|c_{m-1}|^2$. It is important to note that once the inner product information is stored on ancilla, fine threshold functions can be applied. We also note that both parallel and antiparallel $\overline{i}-\overline{w}$ vectors produce perceptron activation, while orthogonal vectors always cause ancilla to be measured in the $|0\rangle_{a}$ state.

This method has been experimentally verified on the IBM quantum computer, and has taken a solid step from theory to practice. However, subject to limited input qubits, the situation with multiple qubits is not clear. In addition, as the number of input bits increases, the demand for quantum gates is getting higher and higher, which may cause unpredictable problems.

In addition to neuron models, a large number of perceptron models have also been proposed in recent years.

\begin{figure}[htbp]
	\vspace{0em}\centering
	\includegraphics[scale=0.5]{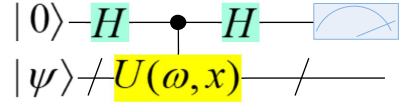}
	\vspace{0em}\caption{The proposed QNN with 1-output and n-input parameters, modified from \cite{50}.  }\label{fig_16}
\end{figure}

\cite{50} proposed a simple QNN with periodic activation function (see Fig. \ref{fig_16}), which only requires $O(n\log_{2}n)$ qubits and $O(nk)$ quantum gates, where \emph{n} is the number of input parameters and \emph{k} is the number of weights applied to these parameters. The corresponding quantum circuit is drawn as follows.

In this circuit with 1-output and \emph{n}-input parameters, $|\psi\rangle$ will be initialized to an equal superposition state, so that the system qubit has an equal effect on the first qubit that produces the output. In the beginning, the initial input of the circuit is determined by
\begin{equation}
	|0\rangle|\psi\rangle=\frac{1}{\sqrt{N}}|0\rangle \sum_{j}^{N}|\mathbf{j}\rangle, \text { with } N=k^{n}
\end{equation} where $|\mathbf{j}\rangle$ is the \emph{j}th vector in the standard basis. Hadamard gate and controlled $U(\omega, x)$ are applied to the first qubit, and the state becomes
\begin{equation}
	\frac{1}{\sqrt{2 N}}(|0\rangle \sum_{i}^{N}|\mathbf{j}\rangle+|1\rangle \sum_{i}^{N} e^{i \alpha_{j}}|\mathbf{j}\rangle)
\end{equation} The parameter $\alpha_{j}$ represents the phase value of the \emph{j}-th eigenvalue of $U$. After passing through the second Hadamard gate, the final state is read as follows:
\begin{equation}
	\frac{1}{\sqrt{2 N}}(|0\rangle \sum_{i}^{N}(1+e^{i \alpha_{j}})|\mathbf{j}\rangle+|1\rangle \sum_{i}^{N}(1-e^{i \alpha_{j}})|\mathbf{j}\rangle)
\end{equation}If the first qubit is measured, the probability of $|0\rangle$ and $|1\rangle$ being $P_{0}$ and $P_{1}$ can be obtained from (18) as
\begin{equation}
	P_{0}=\frac{1}{4 N} \sum_{j}|1+e^{i \alpha_{j}}|^{2}=\frac{1}{2 N} \sum_{j}^{N}(1+\cos (\alpha_{j}))
\end{equation}
\begin{equation}
	P_{1}=\frac{1}{4 N} \sum_{j}|1-e^{i \alpha_{j}}|^{2}=\frac{1}{2 N} \sum_{j}^{N}(1-\cos (\alpha_{j}))
\end{equation}If a threshold function is applied to the output, then
\begin{equation}
	z= \begin{cases}0 & \text { if } P_{1} \leq P_{0} \\ 1 & \text { if } P_{1}>P_{0}\end{cases}
\end{equation}After multiple measurements, \emph{z} is described as the success probability of the expected output, that is, $z=P_{d}$.

The advantage of model of \cite{50} is that a deviation term can be added on this basis to adjust the threshold of the model. In addition, the multi-input generalization ability of the model can have a variety of means, for example, the network can be generalized by sequentially applying  $U_{j}$s.

Since QP is a discrete binary input in many cases, \cite{51} expands the input to a continuous input mode. In order to construct a more complex QP, \cite{52} designs a Multi-Dimensional Input QP (MDIQP) and implemented it by using ancilla qubit input control changes combined with phase estimation and learning algorithms. MDIQP can process quantum information and classify multi-dimensional data that may not be linearly.

\subsection{Others}

\textbf{\emph{QCPNN}} D. Ventura introduces the idea of competitive learning in QNN, and proposes a quantum competitive learning system based on Hamming distance metric \cite{53}. The competitive learning network completes the classification task by comparing the input pattern and the pattern prototype encoded in the network weight. The basic idea is to determine the prototype that is most similar to the input mode (according to some indicators), and the class associated with the prototype is output as the class of the input mode. Based on the classic Hamming neural network algorithm, \cite{54} incorporates quantum theory to obtain a new quantum Hamming neural network based on competitive thinking. This kind of network does not rely on a complete model. Even if the pattern is incomplete, it can still be effectively trained, thereby increasing the probability of pattern recognition. And these unneeded patterns can be further employed as new models for computational training. \cite{55} uses the entanglement measure after the unitary operator to compete between neurons to find the winner category on the basis of winner-takes-all.

\textbf{\emph{QSONN}} SONN is an artificial neural network that adopts an unsupervised competitive learning mechanism. It discovers the internal laws of the input data by adjusting the network parameters and structure through self-organization. \cite{56} earlier proposes a quantum version of SONN, which can perform self-organizing and automatic pattern classification, without the need for a dotted line to store the given pattern, and by modifying the value of the quantum register corresponding to the classification. In order to enhance the clustering ability of QSONN, \cite{57} projects the cluster samples and weights from the competitive layer to the qubits on the Bloch sphere. The winning node can be known by calculating the spherical distance from the sample to the weight. Finally, the samples on the Bloch sphere are updated iteratively according to the weight values of the winning node and its neighborhood until convergence. In addition, following the classic parallel bidirectional self-organizing neural network, \cite{58} proposes its quantum version.

\textbf{\emph{CELL}} \cite{59} introduces the CELL model in 1996. This model is constructed using coupled quantum dot cells in an architecture instead of copying Boolean logic and using physical neighbor connections \cite{59}. In the proposal of \cite{60}, the quantum cellular automaton is regarded as the core neuron cell, and the two-layer quantum cellular automata array forms a three-dimensional CELL which has the structure of A clone template, B clone template and threshold. And the validity of its model is proved in image processing. \cite{61} proposes a fractional-order image encryption CELL model, which uses deformed fractional Fourier transform to solve the problem of insufficient non-linearity. The specific principle is as follows: the input image is processed by the first chaotic random phase mask, and then processed by the first chaotic random phase mask. Finally, the encrypted image is generated in the second chaotic random phase mask as well as the second deformed fractional Fourier transform in sequence. The cryptographic system shows strong resistance to a variety of potential attacks. 

\textbf{\emph{QWLNN}} \cite{62} mentions QWLNN in 2008. \cite{63} defines a QWLNN architecture learning algorithm based on quantum superposition. The architecture and parameters of this model depend on many factors such as the number of training modes, the structure of the selector, etc.

\section{Challenges and Outlook}\label{sec4}

At this stage, although large-scale general-purpose quantum computers have not yet been truly implemented, the recent maturity of quantum processor technology has provided conditions for simple verification of various quantum algorithms. In recent years, benefiting from commercial quantum computers developed by companies such as IBM, researchers can remotely manipulate dozens of qubits through the Internet, build simple quantum circuits, and realize small-scale quantum network systems. On the one hand, it provides a simple experimental verification platform for various QNN models and learning algorithms. On the other hand, it also regulates a strict system framework for QNN theory research, that is, the QNN model and its learning algorithm must be oriented to real quantum circuits and be strictly designed under the quantum mechanics system. In this sense, the research work of QNN still has a long way to go, and the following key scientific issues urgently require further research.

\subsection{Linear and non-linear}

Activation function (such as sigmoid or tanh function), one of core elements in neural networks, has nonlinear characteristics. Its existence makes collective dynamics present dissipative characteristics and attractor-based, and makes it easier for neural networks to capture highly non-trivial patterns \cite{64}-\cite{66}. But this is also the point of divergence from the linear unitary dynamics of quantum computing. Therefore, one core question of QNN is whether it is possible to design a framework to unify the non-linear dynamics of CNN with the unitary dynamics of QNN

In order to solve this problem, the following suggestions can be used for reference: (1) Use simple dissipative quantum gates. (2) Explore the connection between quantum measurements and activation functions. (3) Using quantum circuits to approximate or fit nonlinear functions.

\textbf{\emph{Dissipative quantum gates}} \cite{29} introduces a nonlinear, irreversible, and dissipative operator. This operator can be intuitively regarded as a contraction operator, evolving the general state into a single (stable) state, and the nonlinearity depends only on its amplitude and not on the phase. When designing a QNN, there is an irreversible operator behind the reversible unitary operator. This method has a certain degree of feasibility from a theoretical point of view, but it is very difficult at the level of implementation.

\textbf{\emph{Quantum measurements}} \cite{67} designs a QNN model based on quantum measurement, which attempts to integrate the reversible unitary structure of quantum evolution with the irreversible nonlinear dynamics of neural networks. The author uses an open quantum walk to try to replace the step function or the sigmoid activation function through quantum measurement, and find a quantum form to capture the two main characteristics of the Hopfield network, dissipation and nonlinearity.

\textbf{\emph{Quantum circuits}} The interpretation of nonlinear activation functions by quantum circuits is currently a popular practice. Especially the application of RUS technology to solve the problem of nonlinear activation function \cite{23}\cite{68}-\cite{70}.

\subsection{Verification of quantum superiority}

Limited by the current level of quantum computing hardware, QNN can only perform experiments on low-dimensional and small sample problems, and it is difficult to verify its advantages over CNN. In response to this key issue, it is necessary to establish a unified quantitative index and calculation model to accurately compare the operating complexity and resource requirements of QNN and CNN and to strictly prove the superiority of quantum computing compared to classical computing. In addition, it is necessary to strictly verify the prediction accuracy and generalization performance of the QNN on a large benchmark data set. At present, there are few related studies. \cite{71} and \cite{72} have made an in-depth discussion on the superiority of quantum optimization algorithms for recent quantum processors compared to classical optimization algorithms. Perhaps we can be inspired by them.

\subsection{Barren plateau}

What the Barren Plateau wants to express is that when the amount of qubits is comparatively large, the current QNN framework is easily changed and cannot be effectively trained, that is, the objective function will become very flat, making the gradient difficult to estimate \cite{73}. The root cause of this phenomenon is: according to the objective function constructed by the current quantum circuit (satisfying t-design), the mean value of the gradient of the circuit parameters (some rotation angles) is 0. And the variance exponentially decreases as the total of qubits increases \cite{73}.

\cite{74} extends the Barren Plateau Theorem from a single 2-design circuit to any parameterized quantum circuit, and gives reasonable presumptions so that certain integrals can be expressed as ZX-graphs and calculated using ZX-calculus. The results show that there is a barren plateau for hardware-efficient ansatz and ansatz inspired by MPS, while for QCVNN ansatz and tree tensor network ansatz, there is no barren plateau \cite{74}.

VQA is a commonly useful method of constructing QNN, which optimizes the parameters $\theta$ through the parameterized quantum circuit $V(\theta)$, with the purpose of minimizing the cost function \emph{C}. Considering the connection between it and the barren plateau, \cite{75}  points out that even if $V(\theta)$ is very shallow, defining \emph{C} with a globally observable value will result in a barren plateau. However, as long as the depth of $V(\theta)$ is $O(\log n)$, defining \emph{C} with a locally observable value will lead to a polynomial vanishing gradient in the worst case, thus establishing a connection between locality and trainability.

In order to solve the problem of the barren plateau, it seems to be a good choice to cut from the perspective of initialization. In the scheme proposed by \cite{76}, the first step is to randomly select some initial parameter values, and then select the remaining ones. Such circuits constitute a sequence of shallow blocks, and each shallow block calculates the identity, which controls the effective depth of the circuit for a parameter update, so that they will not enter the barren plateau at the beginning of training.

The above references are only a useful attempt on the barren plateau, but the problem of the barren plateau has not been solved perfectly and is a problem worthy of study.

\section*{Acknowledgment}
We would like to thank all the reviewers who provided valuable suggestions and Chen Zhaoyun, Ph.D., Department of Physics, University of Science and Technology of China.

%The authors would like to thank the students who took the Digital Systems II course and whose suggestions and responses were used here. The authors would like to thank the Mechatronics Laboratory, the Autonomous University of Quer¨¦taro, for granting the permission to use their facilities to implement this project. The first author would like to thank CONACyT for his Postdoctoral Fellowship.
\ifCLASSOPTIONcaptionsoff
  \newpage
\fi

\end{document}